# Downburst Prediction Applications of Meteorological Geostationary Satellites

Kenneth L. Pryor*

National Oceanic and Atmospheric Administration (United States)/Center for Satellite Applications and Research, 5830 University Research Court, College Park, MD 20740

**ABSTRACT**

A suite of products has been developed and evaluated to assess hazards presented by convective storm downbursts derived from the current generation of Geostationary Operational Environmental Satellite (GOES) (13-15). The existing suite of GOES downburst prediction products employs the GOES sounder to calculate risk based on conceptual models of favorable environmental profiles for convective downburst generation. A diagnostic nowcasting product, the Microburst Windspeed Potential Index (MWPI), is designed to infer attributes of a favorable downburst environment: 1) the presence of large convective available potential energy (CAPE), and 2) the presence of a surface-based or elevated mixed layer with a steep temperature lapse rate and vertical relative humidity gradient. These conditions foster intense convective downdrafts upon the interaction of sub-saturated air in the elevated or sub-cloud mixed layer with the storm precipitation core. This paper provides an updated assessment of the MWPI algorithm, presents recent case studies demonstrating effective operational use of the MWPI product over the Atlantic coastal region, and presents validation results for the United States Great Plains and Mid-Atlantic coastal region. In addition, an application of the brightness temperature difference (BTD) between GOES imager water vapor (6.5μm) and thermal infrared (11μm) channels that identifies regions where downbursts are likely to develop, due to mid-tropospheric dry air entrainment, will be outlined.

## 1. INTRODUCTION

A suite of products has been developed and evaluated to assess hazards presented by convective-storm generated downbursts[1] derived from the current generation of Geostationary Operational Environmental Satellite (GOES-13 to 15). The existing suite of GOES sounder[2] derived microburst products are designed to diagnose risk based on conceptual models of favorable environmental profiles for severe convective storm development. The typical size scale of a single-cell convective storm[3] and larger downburst (or macroburst) is near 10 kilometers (km). Considering the 10-km spacing of sounding retrievals, the GOES sounder is well suited to observe environmental conditions and associated parameters that indicate downburst risk. The development of a GOES sounder-derived wet microburst severity index (WMSI) product[4] has been outlined to calculate the potential magnitude of convective downbursts in humid environments over the eastern United States. The WMSI incorporated convective available potential energy (CAPE) as well as the vertical theta-e difference[5] ($\Delta\theta_e$) between the surface and mid-troposphere. Knupp[6], through numerical simulations, confirmed that entrainment of dry mid-level air into the downshear flank of a convective storm fostered strong downdraft generation. A saturation mixing point analysis detailed in this study found that the entrainment of subsaturated air into the downshear flank of a convective storm results in a significant wet bulb potential temperature depression that is proportional to the cooling due to evaporation of precipitation. The negative buoyancy induced by this mid-level cooling forces downdraft initiation and acceleration within a deep, moist convective storm.

The majority of microburst days during Joint Airport Weather Studies (JAWS) project conducted in the Denver, Colorado area were characterized by environments intermediate between the dry and wet extremes[7] (i.e. "hybrid"). In a prototypical dry microburst environment, Wakimoto[8] identified a convective cloud base height near 500 mb associated with an "inverted V" profile. In contrast, Atkins and Wakimoto[5] identified that a typical cloud base height in a wet microburst environment is near the 850-mb level. Thus, a cloud base height of 670 mb was proposed for a hypothetical, weak shear intermediate, or "hybrid" microburst environment. Pryor[9] outlined the selection process for the upper boundary level of 670-mb for a microburst wind speed potential calculation.

*Ken.Pryor@noaa.gov; 1 301 683-3575

Johns and Doswell[10] identified necessary ingredients for deep convection. CAPE has an important role in precipitation formation due to the strong dependence of updraft strength and resultant storm precipitation content on positive buoyant energy. Loading of precipitation, typically in the form of graupel and hail, initiates the convective downdraft. The subsequent melting of graupel and hail and sub-cloud evaporation of liquid precipitation result in cooling and the negative buoyancy that accelerate the downdraft in the unsaturated layer. Collectively, melting of graupel, subsequent evaporative cooling and resulting downdraft strength are enhanced by large liquid water content and related water surface available for evaporation, and a large lapse rate that acts to maintain negative buoyancy as the downdraft descends in the sub-cloud layer. The intense downdraft subsequently produces strong and potentially damaging winds upon impinging on the surface. Ellrod[11] noted that lapse rates, specifically between 700 and 850 mb, contained the most predictive information for determining downburst potential using GOES sounder profile data. In addition, it has been found that afternoon sub-cloud temperature lapse rates were strongly correlated with microburst activity[12].

The Microburst Windspeed Potential Index (MWPI) is designed to quantify the most relevant factors in convective downburst generation in intermediate thermodynamic environments by incorporating: 1) CAPE, 2) temperature lapse rate between the 670 and 850-mb levels ($\Gamma$), and 3) the dew point depression difference ($\Delta DD$) between the 670-mb and 850-mb levels. The MWPI is incorporated into a predictive linear model developed in the manner exemplified by Caracena and Flueck[7]. The MWPI formula consists of a set of predictor variables (i.e. dewpoint depression, temperature lapse rate) that generates output of expected microburst risk. Thus, the MWPI algorithm that accounts for both updraft and downdraft instability in microburst generation is defined as

$$MWPI \equiv \{(CAPE/100)\} + \{\Gamma + (T-T_d)_{850} - (T-T_d)_{670}\} \qquad (1)$$

where $\Gamma$ is the lapse rate (°K km$^{-1}$) between the lower boundary level (850 mb) and upper boundary level (670 mb) of a layer of consideration, and the quantity $(T-T_d)$ represents the dewpoint depression (°K). The MWPI algorithm is expected to be most effective in assessing downburst wind gust potential associated with "pulse"-type (short duration, single cell) convective storms in weak wind shear environments.

Generation of the MWPI product is based on the following assumptions: 1) mixed-phase precipitation, in the form of graupel, hail, and supercooled rain, is present in the middle level of the storm, 2) phase-change cooling (sublimation, melting, and evaporation) is the primary forcing factor in negative buoyancy generation and subsequent acceleration of convective storm downdrafts, 3) precipitation loading is a secondary forcing mechanism, and 4) the freezing level and the level of minimum equivalent potential temperature are located at or above the 670-mb level. Although the MWPI algorithm was originally designed for convective wind speed potential assessment in intermediate thermodynamic environments over the central United States, the MWPI has demonstrated effectiveness for both wet microbursts that occur over the eastern U.S. as well as for dry microbursts that occur over the western U.S. intermountain region. For wet microburst environments, large CAPE and conditionally unstable temperature lapse rates (5 to 10 °K km$^{-1}$) between the 670 and 850-mb levels would be readily detected by the MWPI algorithm and indicate the potential for intense deep moist convection. Precipitation loading, melting of graupel and hail, and evaporation of liquid precipitation in the layer between 670 and 850 mb would enhance negative buoyancy and downdraft acceleration in wet-microburst producing convective storms.

The temperature lapse rate and dewpoint depression difference terms effectively indicate a favorable lower tropospheric thermodynamic structure for microbursts in a low CAPE environment. Further statistical analysis of a dataset built by comparing wind gust speeds recorded by Oklahoma Mesonet stations to adjacent MWPI values for 35 downburst events has yielded some favorable results, as displayed in Figure 1. Correlation was computed between key parameters in the downburst process, including temperature lapse rate ($\Gamma$) and radar reflectivity. The first important finding is a statistically significant negative correlation (r=-.34) between lapse rate and radar reflectivity. Similar to the findings of Srivastava[13], for lapse rates greater than 8 °K km$^{-1}$, downburst occurrence is nearly independent of radar reflectivity. For lapse rates less than 8 °K km$^{-1}$, downburst occurrence was associated with high reflectivity (> 50 dBZ) storms. The majority of downbursts occurred in sub-cloud environments with lapse rates greater than 8.5 °K km$^{-1}$.

In addition, it has been found recently that the brightness temperature difference ("BTD") between GOES infrared band 3 (water vapor, 6.5μm) and band 4 (thermal infrared, 11μm) can highlight regions where downburst generation is likely due to the channeling of unsaturated mid-tropospheric air into the precipitation core of a deep, moist convective storm. Rabin et al.[14] noted that observations have shown that BTD > 0 can occur when water vapor exists above cloud tops in a stably stratified lower stratosphere and thus, BTD > 0 has been used a measure for intensity of overshooting convection. A new feature presented in this paper readily apparent in BTD imagery is a "dry-air notch", identified in GOES infrared imagery as a V- or U-shaped region of relatively warm brightness temperature that typically appears on the rear flank of downburst-producing convective storms. The dry-air notch can signify the development of a downburst as unsaturated air is being channeled into the storm precipitation core. Important in the downburst generation process, as indicated by the presence of a dry-air notch, is the mid-tropospheric value of $\theta_e$ and its vertical profile.

## 2. DATA COLLECTION AND METHODOLOGY

The main objective of the validation effort is to qualitatively and quantitatively assess the performance of the MWPI algorithm by employing classical statistical analysis of real-time data. Accordingly, this effort entails a study of downburst events in a manner that emulates historic field projects such as the 1982 Joint Airport Weather Studies (JAWS)[8] and the 1986 Microburst and Severe Thunderstorm (MIST) project[5]. Algorithm output data was collected for downburst events that occurred during the warm season (especially between 1 June and 30 September) and was validated against surface observations of convective wind gusts as recorded by wind sensors in high-quality mesonetworks, such as the Oklahoma[15] and West Texas Mesonets[16], and over the Chesapeake Bay region by NOAA marine network stations. Figure 2 compares the geographic regions of interest within the continental United States (CONUS) for the validation of the MWPI algorithm. Note that the Oklahoma-Texas region, with an area near $4 \times 10^5$ km$^2$, is about 10 times larger than the Chesapeake Bay region that is sampled for this validation study. The 10-km spacing of GOES sounding retrievals and resulting microburst risk algorithm output in clear-sky regions plotted over a visible or infrared GOES image facilitates the co-location of index values and measured downburst-related wind speeds at the surface. In effect, the horizontal resolution of the GOES sounder ensures the representativeness of MWPI values that are in close proximity to observations of downburst winds. Wakimoto[8] and Atkins and Wakimoto[5] discussed the effectiveness of using mesonetwork surface observations and radar reflectivity data in the verification of the occurrence of downbursts.

As illustrated in the flowchart in Figure 3, real-time experimental MWPI product images are generated by Man computer Interactive Data Access System (McIDAS)-X. This program reads and processes GOES sounder data, calculates and collates microburst risk values, and overlays risk values on GOES imagery. Output images are then archived via FTP and HTTP to the GOES Microburst Products web page. For selected downburst events, the MWPI product was generated using McIDAS-V*. The MWPI algorithm, as visualized by McIDAS-V, reads and processes GOES sounder profile data in binary format. The MWPI is then calculated for each retrieval location and plotted on a user-defined map.

The BTD product consisted of image data derived brightness temperatures from the GOES-East and West imager 4-km resolution water vapor (band 3) and thermal infrared (band 4), obtained from the Comprehensive Large Array-data Stewardship System (CLASS). BTD algorithm output was visualized by the NOAA Weather and Climate Toolkit. A data stretch and built-in color enhancement were applied to the output images to highlight patterns of interest including overshooting tops and dry-air notches. The dry-air notch identified in the BTD image is analogous in concept to the rear-inflow notch (RIN) as identified in radar imagery[17]. Next Generation Radar (NEXRAD) and Terminal Doppler Weather Radar (TDWR) base reflectivity imagery from National Climatic Data Center (NCDC) were utilized to verify that observed wind gusts are associated with high-reflectivity downbursts and not associated with other types of convective wind phenomena (i.e. gust fronts). An application of radar imagery is to infer microscale physical properties of downburst-producing convective storms. Particular radar reflectivity signatures, such as the RIN[17], the spearhead echo[18], and protrusion echo[19] are effective indicators of the occurrence of downbursts.

Since surface data quality is paramount in an effective validation program, relatively flat, treeless prairie regions were initially chosen as study regions. The treeless, low-relief topography that dominates sparsely populated regions such as the U.S. High Plains allowed for the assumption of horizontal homogeneity when deriving a conceptual model of a boundary layer thermodynamic structure favorable for downbursts. More importantly, planar topography and water body surfaces facilitate relatively smooth flow (due to small surface roughness) with respect to downburst winds in which drag and turbulent eddy circulation resulting from surface obstructions (i.e., buildings, hills, and trees) are minimized. Over the eastern United States, Atlantic coastal waters are optimal for a validation study. Observational data from Integrated Ocean Observing System (IOOS) stations and WeatherFlow* was preferred for algorithm validation. It is also assumed that there is very little change in physical characteristics of storms that move from land areas over the Chesapeake Bay.

* Available online at http://www.ssec.wisc.edu/mcidas/software/v/

In order to assess the predictive value of the algorithm output, the closest representative index values were obtained for retrieval times one to three hours prior to the observed surface wind gusts. Representativeness of proximate index values was ensured by determining from analysis of surface observations, radar, and satellite imagery that no change in environmental static stability and air mass characteristics between product valid time and time of observed downbursts had occurred. Furthermore, in order for the downburst observation to be included in a validation data set, it was required that the parent convective storm cell of each downburst, with radar reflectivity greater than 35 dBZ, be located nearly overhead at the time of downburst occurrence. An additional criterion for inclusion into a data set is a wind gust measurement of at least 18 m s$^{-1}$ (35 kt) that is widely considered to be operationally significant for transportation, especially boating and aviation. A technique[8] to visually inspect wind speed observations over the time intervals encompassing candidate downburst events was implemented to exclude gust front events from the validation data set. In summary, the screening process employed to build the validation data set that consists of criteria based on surface weather observations and radar reflectivity data yielded a statistically significant sample size of downbursts and associated index values.

Similar to previously developed GOES-derived microburst products, it is important to emphasize that the wind gust potential expressed in the MWPI is conditional on the occurrence of convective storms and thus, validation metrics such as probability of detection (POD) and false alarm ratio (FAR) were not considered to be appropriate for this study. Thus, covariance and mean difference between the variables of interest, MWPI and surface downburst wind gust speed, were analyzed. Algorithm effectiveness was assessed as the correlation (r) between MWPI values and observed surface wind gust velocities, as well as the mean absolute error (MAE) between MWPI-predicted downburst wind gust speeds (based on the linear regression equations noted in Table 1) and observed surface wind gust speeds. Statistical significance testing, specifically, a t-test for correlated samples[20] was conducted to determine the confidence level of the correlation between observed downburst wind gust magnitude and microburst risk values. Examples of MWPI algorithm validation employing the direct comparison method are shown graphically in the case studies in Section 4. It is also important to emphasize that MWPI values are not expressed in dimensions of wind speed, but are related to wind speed through the regression line equations that appear in Table 1.

## 3. VALIDATION RESULTS

Product validation for the MWPI product was conducted for two distinct regions of the continental United States: The southern Great Plains region between 2007 and 2010 and the Mid-Atlantic coastal region between 2010 and 2013. Table 1 outlines the results of validation statistical analysis over these two distinct climatic regions: the southern Great Plains region varies from semi-arid to humid from west to east, while the Mid-Atlantic region is entirely humid. The mean absolute error (MEA) represents the average difference between the MWPI-derived predicted wind gust speed ("y") based on regression line equations for each region, and the observed downburst wind gust speed.

Over the southern Plains region, GOES sounder-derived MWPI values were directly compared to mesonet observations of downburst winds over Oklahoma and Texas for 208 events between June 2007 and September 2010. The correlation (r) between MWPI values and measured wind gusts was determined to be 0.62 and was found to be statistically significant near the 100% confidence level, indicating that the correlation represents a physical relationship between MWPI values and downburst magnitude and is not an artifact of the sampling process.

Within the Mid-Atlantic coastal region, the Chesapeake Bay area was the region of focus for the validation study. The Maryland portion of the Chesapeake Bay watershed is located within an area with an elevated frequency of severe thunderstorm wind occurrence[21]. In addition, a dense marine weather observing network including NOAA tower-mounted stations and weather data buoys that exists over the Chesapeake Bay and adjacent estuaries allowed for frequent sampling of downburst-related winds. Direct comparison between MWPI values and observed wind gusts from the period of summer 2010 to summer 2013 yielded 87 recorded downburst wind events. Based on this statistically significant data sample, a correlation between MWPI values and measured wind gusts of 0.74 was found for this study. Interestingly, this correlation is higher than that found for the southern Great Plains, the region for which the MWPI was originally developed. In addition, a mean absolute error (MEA) of slightly less than one knot (0.96) between MWPI-derived winds and measured winds was found over the Chesapeake Bay region. For both regions, *t*-test results, in which computed *t* values were much larger than the critical values associated with each respective sample size indicated that correlation between MWPI values and adjacent measured downburst wind gust speeds were statistically significant beyond the 99.95% confidence level.

*http://datascope.weatherflow.com

## 4. CASE STUDY: MARYLAND CHESAPEAKE BAY DOWNBURSTS

During the late afternoon and evening of 10 July 2013, multicellular convective storms developed over the Virginia piedmont region and then tracked east and northeastward toward the Maryland western shore of the Chesapeake Bay. These multicell storms developed and evolved in a warm, moist and generally unstable air mass well ahead of a decaying mesoscale convective system (MCS) moving into the western Appalachian Mountains and a cold front moving through the Ohio Valley region. During the early evening, between 2300 UTC 10 July 2013 and 0000 UTC 11 July 2013, a particularly intense multicell storm developed over the Maryland Chesapeake Bay western shore and tracked northeastward over the upper Chesapeake Bay between Annapolis and Baltimore. Figure 4, late afternoon (2100 and 2200 UTC 10 July) GOES MWPI product images, illustrates the development of deep convective storms over the Maryland-Virginia region where elevated (yellow) values indicated wind gust potential of 18 to 25 m s$^{-1}$ (35 to 49 kt). Figure 5 shows a closer view of GOES-13 sounder derived MWPI and $\Delta\theta_e$ parameters near 2100 UTC. Note a general decrease in MWPI values northward over the Chesapeake Bay, while $\Delta\theta_e$ values were uniformly high (> 20°K) over and west of the bay. The microburst product image shows a local maximum in values over the western shore of the Chesapeake Bay during the late afternoon. Application of the linear regression technique to MWPI values of 30 to 42 (white-circled region) that were the closest representative values to the Annapolis area yielded wind gust potential of 22 to 25 m s$^{-1}$ (42 to 48 kt) over three hours prior to downburst wind occurrence. In a similar manner, an MWPI value of 17.6 near the upper Chesapeake Bay was associated with wind gust potential near 19 m s$^{-1}$ (37 kt). As documented in Table 2, marine weather observing stations in the Annapolis area recorded downburst wind gusts between 22 and 25 m s$^{-1}$ (42 and 48 kt) between 0000 and 0040 UTC 11 July 2013 as an intense cell on the leading edge of the storm tracked over Annapolis Harbor. Figure 6, a GOES-derived sounding profile over Baltimore at 2200 UTC graphically describes attributes of the pre-convective environment that indicated potential for intense storm downdrafts and resulting strong downburst winds. Of interest is a layer in the lower to mid-troposphere, between the 650 and 850-mb levels, with a conditionally unstable temperature lapse rate and increasing dew point depression with decreasing height above ground level. This representative sounding profile supports the ability of the MWPI algorithm to detect a statically unstable environment expressed as elevated index values with downburst wind gust potential greater than 18 m s$^{-1}$ (35 kt). Figure 7 confirms downburst occurrence as inferred from the divergent nature of the winds observed at Greenbury Point WeatherFlow station and the NOAA Annapolis buoy near 0030 UTC 11 July 2013 with respect to Baltimore, Maryland TDWR reflectivity imagery. TDWR imagery in Figure 7a displayed protrusion echoes and a RIN associated with the downburst-producing storm. Near the time of downburst occurrence over the Annapolis Harbor, GOES WV-IR BTD imagery in Figure 7b indicated the presence of both rear and forward flank mid-tropospheric dry-air notches associated with the multicell storm tracking over the Annapolis Harbor. Video 1, a GOES BTD/TDWR reflectivity composite image animation, shows the evolution of the multicell storm and the northward expansion of a dry-air notch on the southeastern flank during the time of greatest downburst intensity.

After 0030 UTC, the multicell storm continued to track northward over the upper Chesapeake Bay and into a slightly more stable environment as indicated by an MWPI values between 20 and 30. Between 0050 and 0100 UTC, the NOAA Patapsco Buoy near Baltimore, recorded a weaker downburst wind gust of 19 m s$^{-1}$ (37 kt): exactly the value predicted by the MWPI regression equation. In addition, the MWPI product was compared to an existing weather radar algorithm[22] that employs Doppler radar-measured echo tops (ET) and vertical integrated liquid (VIL) products to calculate downburst wind gust potential (WGP). The results of this comparison study are also noted in Table 2. Based on this statistically significant data sample, a correlation between MWPI values and measured wind gusts of 0.74 was found for this downburst event. The correlation of the uncorrected ET/VIL-derived WGP was considerably lower (r=0.18) while mean error for the ET/VIL WGP was much higher (-10 kt) as compared to the mean error of the MWPI WGP (0.1 kt). This case demonstrated the capability of the GOES MWPI product to effectively indicate wind gust potential with a significantly longer lead-time as compared to the Doppler radar technique. Overall, the MWPI product effectively indicated the presence of strong instability and resulting downburst potential over the Chesapeake Bay region of Maryland over three hours prior to the observance of high winds. Interestingly, no special marine warnings were issued or any local storm reports documented for the convective winds measured between 19 and 24 m s$^{-1}$ (36 and 46 kt) by WeatherFlow stations and NOAA data buoys located on the upper Chesapeake Bay.

## 5. DISCUSSION AND CONCLUSIONS

Pryor[9] presents case studies of significant downburst events that occurred over the southern Great Plains. The case study presented in this paper highlights the adaptability and flexibility of the MWPI algorithm to use in varying climatic and geographic regions in the continental United States. Thus, the MWPI algorithm that can be refined or "tuned" based on regional climatology.

In general, MWPI values of 50 or greater correspond to convective wind gust potential of 26 m s$^{-1}$ (50 kt) or greater. This threshold is important for the issuance of severe thunderstorm warnings by the National Weather Service. Case studies over the Great Plains and Mid-Atlantic region[9] have demonstrated the ability of the MWPI algorithm to accurately predict wind gust potential over 26 m s$^{-1}$ (50 kt). The downburst event discussed in this paper, over the Chesapeake Bay region, highlights the strength of the MWPI algorithm to anticipate downburst wind events that can be operationally significant for boating and aviation (18 to 25 m s$^{-1}$ or 34 to 49 kt) but do not meet the criteria for a severe thunderstorm warning. Sustained thunderstorm winds or associated gusts of 18 m s$^{-1}$ (34 kt) or greater over coastal and ocean waters and major estuaries (such as the Chesapeake Bay) do meet the criteria for issuance of a special marine warning. The MWPI has demonstrated the capability to forecast, with up to four hours lead time, thunderstorm-generated wind gusts that meet special marine warning criteria. In addition, the most intense downburst occurrence was found near local maxima in MWPI values, as highlighted in Figure 5. Although during the Chesapeake Bay downburst event, the $\Delta\theta_e$ parameter correlated poorly with downburst wind gust magnitude, $\Delta\theta_e$ values were consistently greater than 20 °K. $\Delta\theta_e$ values greater than 20 have been previously associated with a high likelihood of wet-type downburst occurrence and has functioned as a supplement to the MWPI algorithm in forecasting downburst wind gust potential.

The dry-air notch presented above likely represents unsaturated air that is entrained into convective storms and interacts with their precipitation cores, subsequently providing the energy for intense downdrafts and resulting downburst winds. Comparison of BTD product imagery to corresponding radar imagery revealed physical relationships between the dry-air notch, rear-inflow notch and the spearhead (or protrusion) echo. Entrainment of drier mid-tropospheric air into the precipitation core of the convective storm typically results in evaporation of precipitation, the subsequent cooling and generation of negative buoyancy, and resultant acceleration of a downdraft. When the intense localized downdraft reaches the surface, air flows outward as a downburst. Thus, the WV-IR band BTD product can serve as an effective supplement to the GOES sounder MWPI product, especially in regions where there is no Doppler radar coverage (i.e. over open ocean waters).

As proven by statistical analysis, the GOES sounder MWPI product has demonstrated capability in the assessment of wind gust potential over the southern Great Plains and Mid-Atlantic coast regions. Statistical analysis for downburst events that occurred during the 2007 to 2013 convective seasons and a recent case study from the 2013 convective season demonstrated the effectiveness of the GOES MWPI algorithm as evidenced by a statistically significant correlation between MWPI values and measured downburst wind gusts and a low mean error, less than one knot, between predicted wind gust speeds derived from the MWPI regression equation and observed wind gust speeds. Further validation over the Atlantic coast region has served to strengthen the functional relationship between MWPI values and downburst wind gust magnitude as well as demonstrate the adaptability of the MWPI algorithm to diverse climatic and geographic regions.

Additional validation over geographically diverse regions in the continental United States such as Florida, the Great Lakes, and the intermountain western U.S. and quantitative statistical analysis to assess product performance will serve as future work in the development and evolution of the GOES MWPI product.

## ACKNOWLEDGEMENTS

The author thanks the Oklahoma and West Texas Mesonets, and WeatherFlow, Inc. for the surface weather observation data used in this research effort. The author also thanks Jaime Daniels (NESDIS) for providing GOES sounding retrievals displayed in this paper.

# TABLES AND FIGURES

Table 1. MWPI validation statistics based on direct comparison between index values and measured downburst wind gusts. In the regression line equations, "x" represents the MWPI value while "y" represents predicted wind gust speed.

|  | Oklahoma-Texas (N=208) | Mid-Atlantic (N=87) |
|---|---|---|
| MEA (kt) | -0.55 | 0.96 |
| Correlation (r) | 0.62 | 0.74 |
| *t* value | 9.23 | 11.78 |
| Critical Value (P< 0.0005) | 3.34 | 3.41 |
| Regression line equation | y=0.3163x+33.766 | y=0.4553x+28.769 |

Table 2. Measured wind gusts and associated microburst risk values for the 11 July 2013 Maryland Chesapeake Bay downburst event. WeatherFlow stations are identified by "WF".

| Station | Time (UTC) | Wind Gust Speed (kt) | MWPI | $\Delta\theta_e$ (K) | MWPI WGP (kt) | ET-VIL WGP (kt) |
|---|---|---|---|---|---|---|
| Tolly Point (WF) | 0022 | 39 | 31 | 21 | 43 | 34 |
| Greenbury Point (WF) | 0033 | 48 | 31 | 21 | 43 | 33 |
| Annapolis Buoy (NDBC) | 0040 | 42 | 31 | 21 | 43 | 27 |
| Patapsco Buoy (NDBC) | 0100 | 37 | 18 | 24 | 37 | 31 |

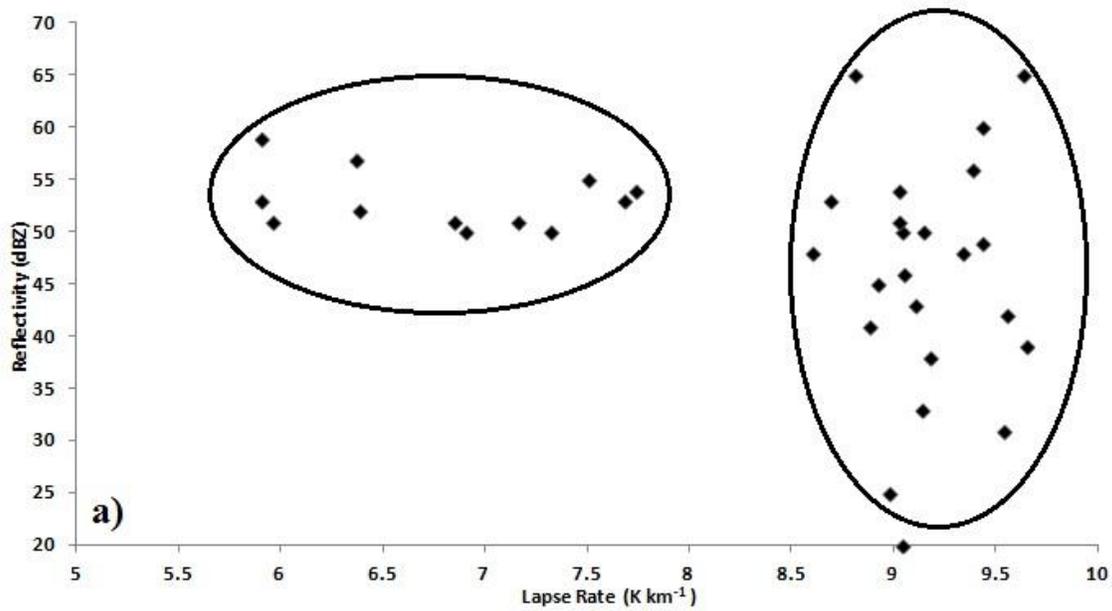

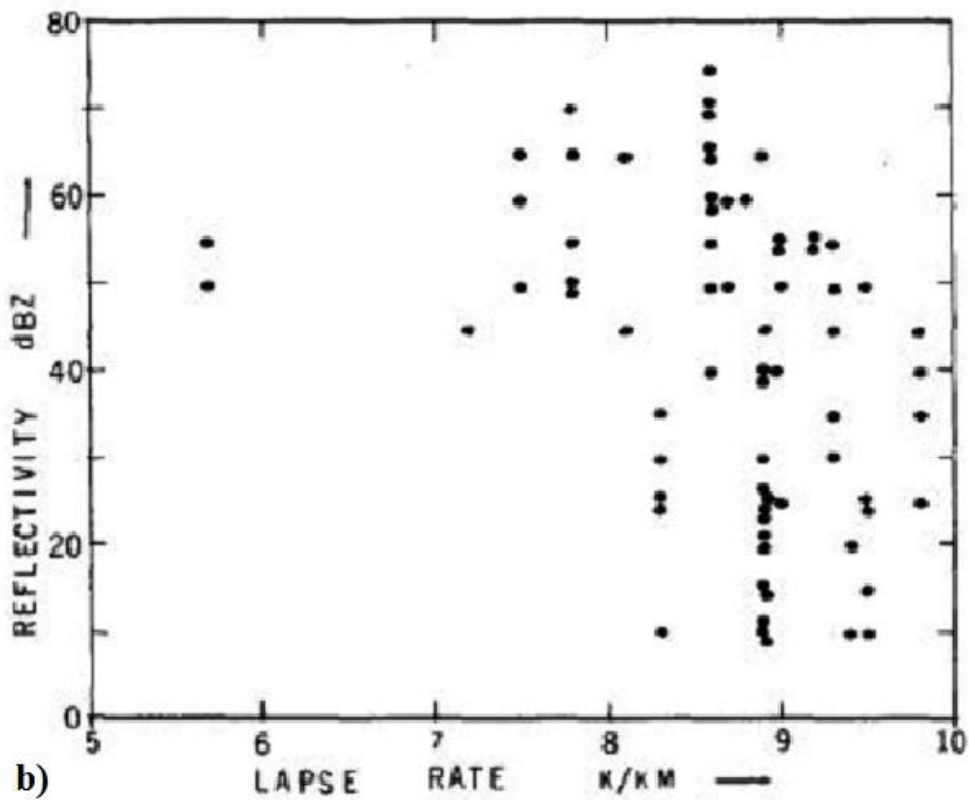

Figure 1. a) Scatterplot of lapse rate versus radar reflectivity for 35 downburst events over Oklahoma during the 2009 convective season compared to b) scatterplot for 186 microburst events during the 1982 JAWS project (courtesy Srivastava 1985).

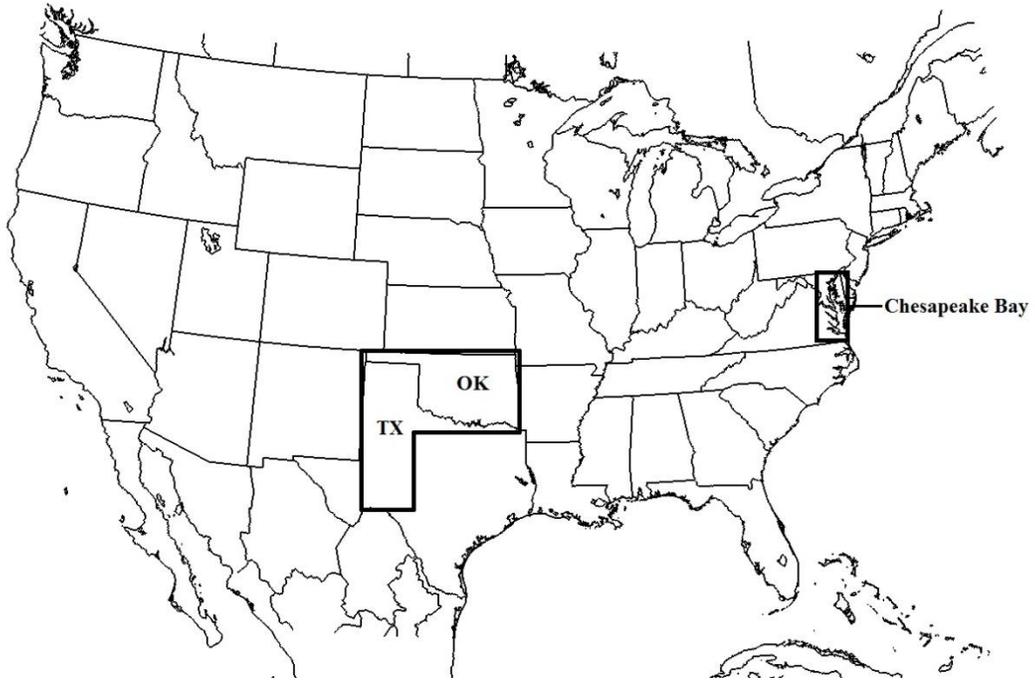

Figure 2. Geographic regions of interest within the continental United States (CONUS) for the validation of the MWPI algorithm.

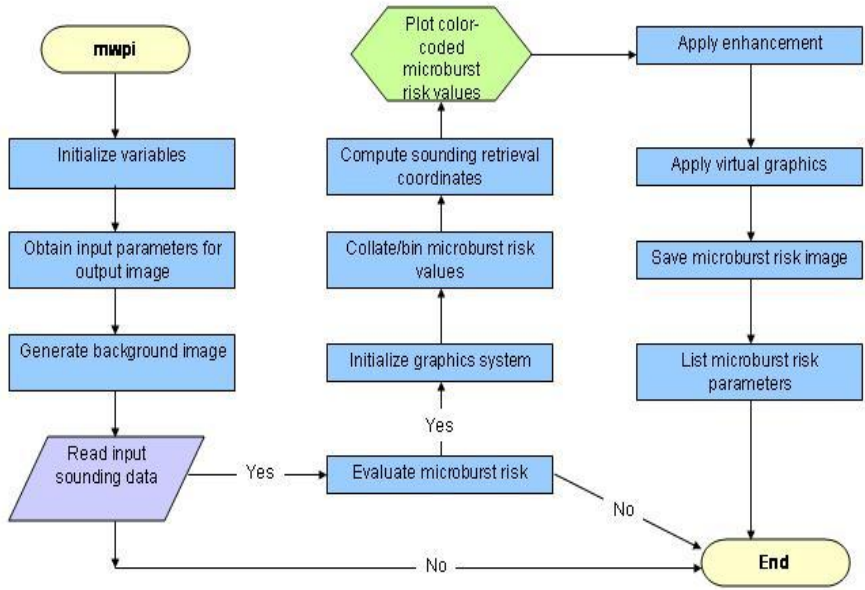

Figure 3. Flowchart illustrating the operation of the MWPI program in the McIDAS-X environment.

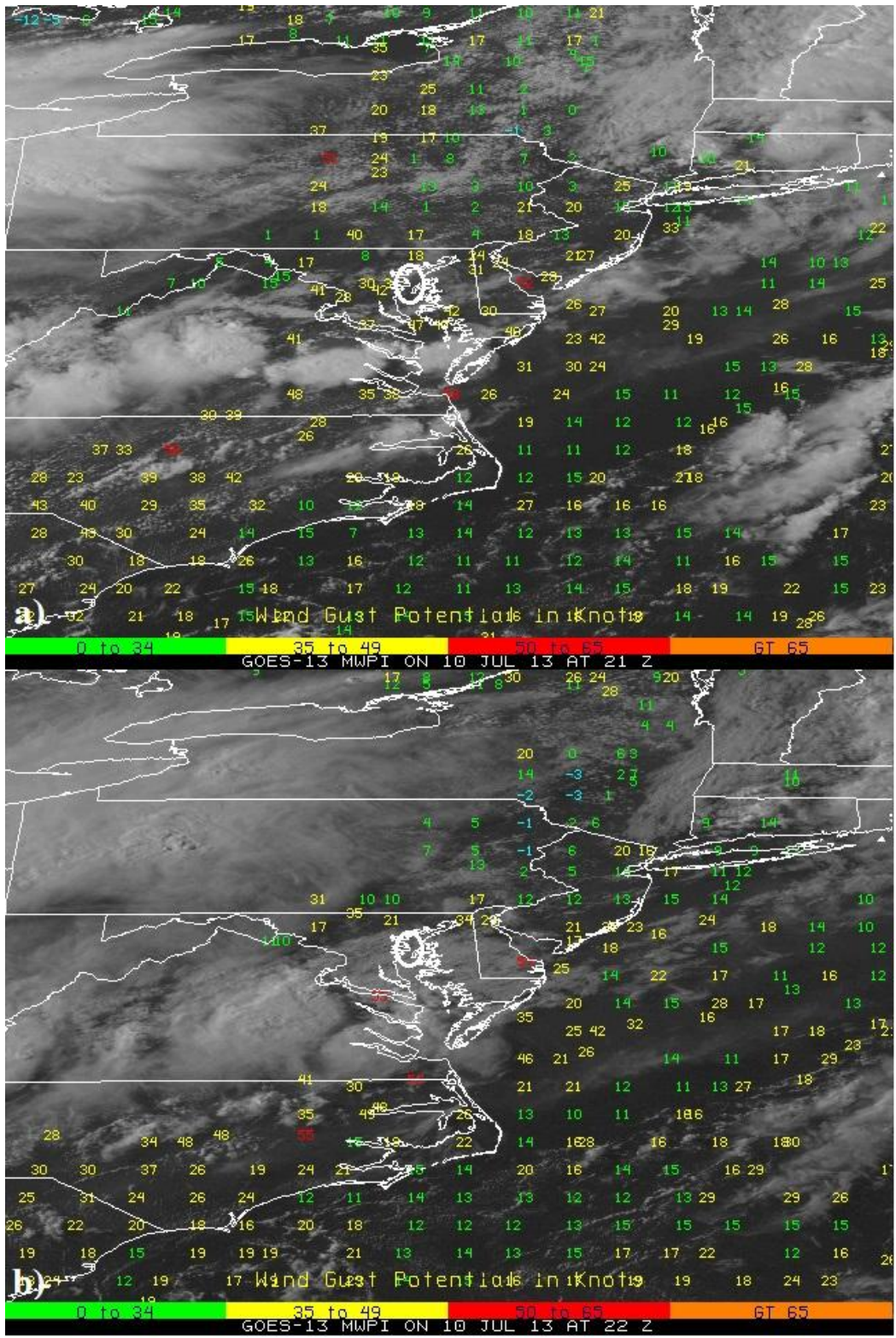

Figure 4. Mid-Atlantic sector GOES MWPI product images at a) 2100 UTC and b) 2200 UTC 10 July 2013. White-circled region outlines area of downburst activity during the evening of 10 July.

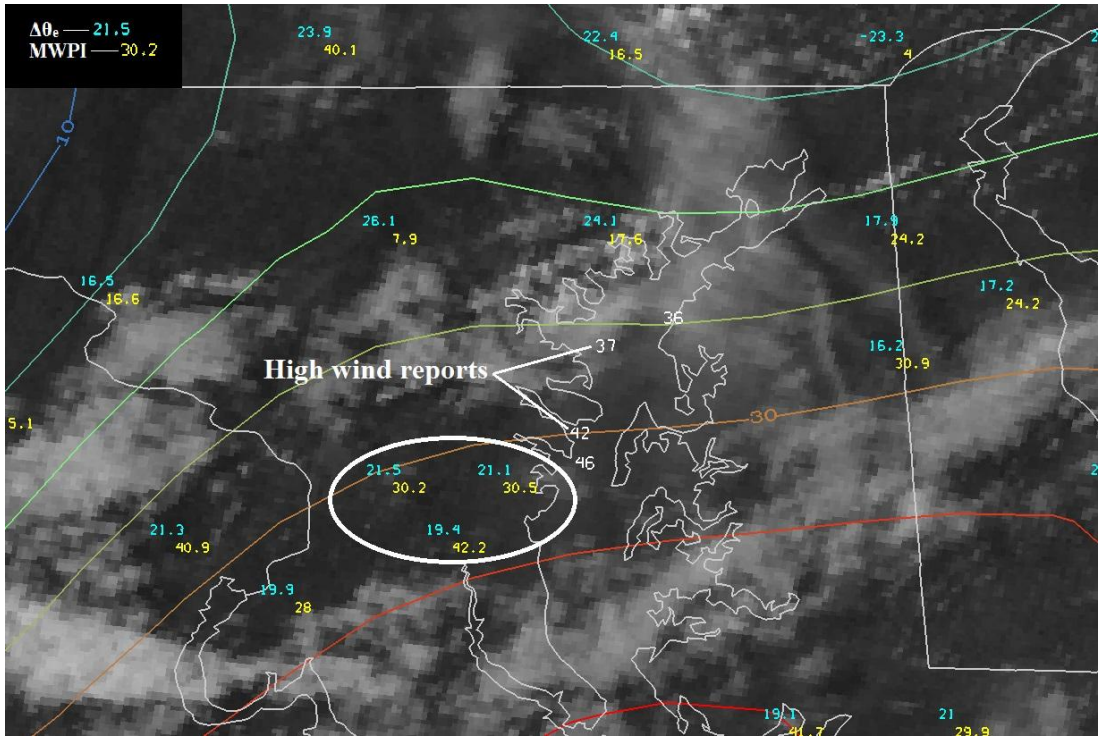

Figure 5. Local scale GOES-13 microburst product image over the Maryland Chesapeake Bay region at 2045 UTC 10 July 2013. Colored contours of MWPI values are overlying MWPI values (plotted in yellow type), Δθ$_e$ values (plotted in blue type), and a GOES-13 visible image. Contour interval is 5, with contours representing values of 10 and 30 labeled. Local maximum in MWPI values is highlighted by a white-circled region.

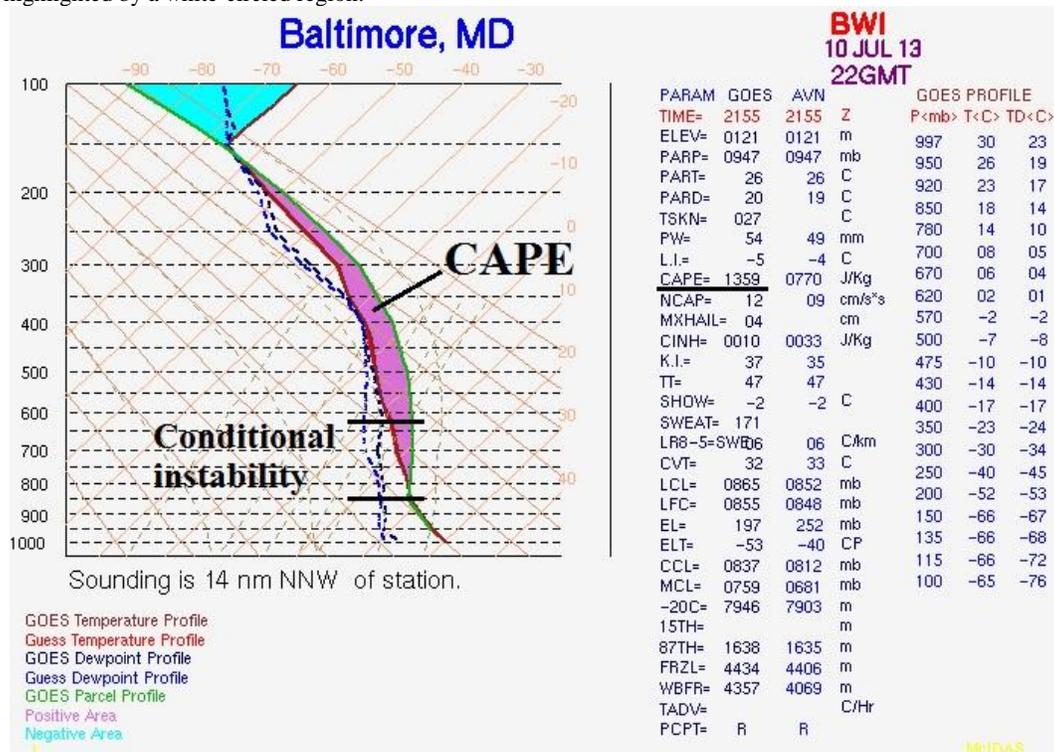

Figure 6. GOES sounding profile over Baltimore, Maryland at 2200 UTC 10 July 2013.

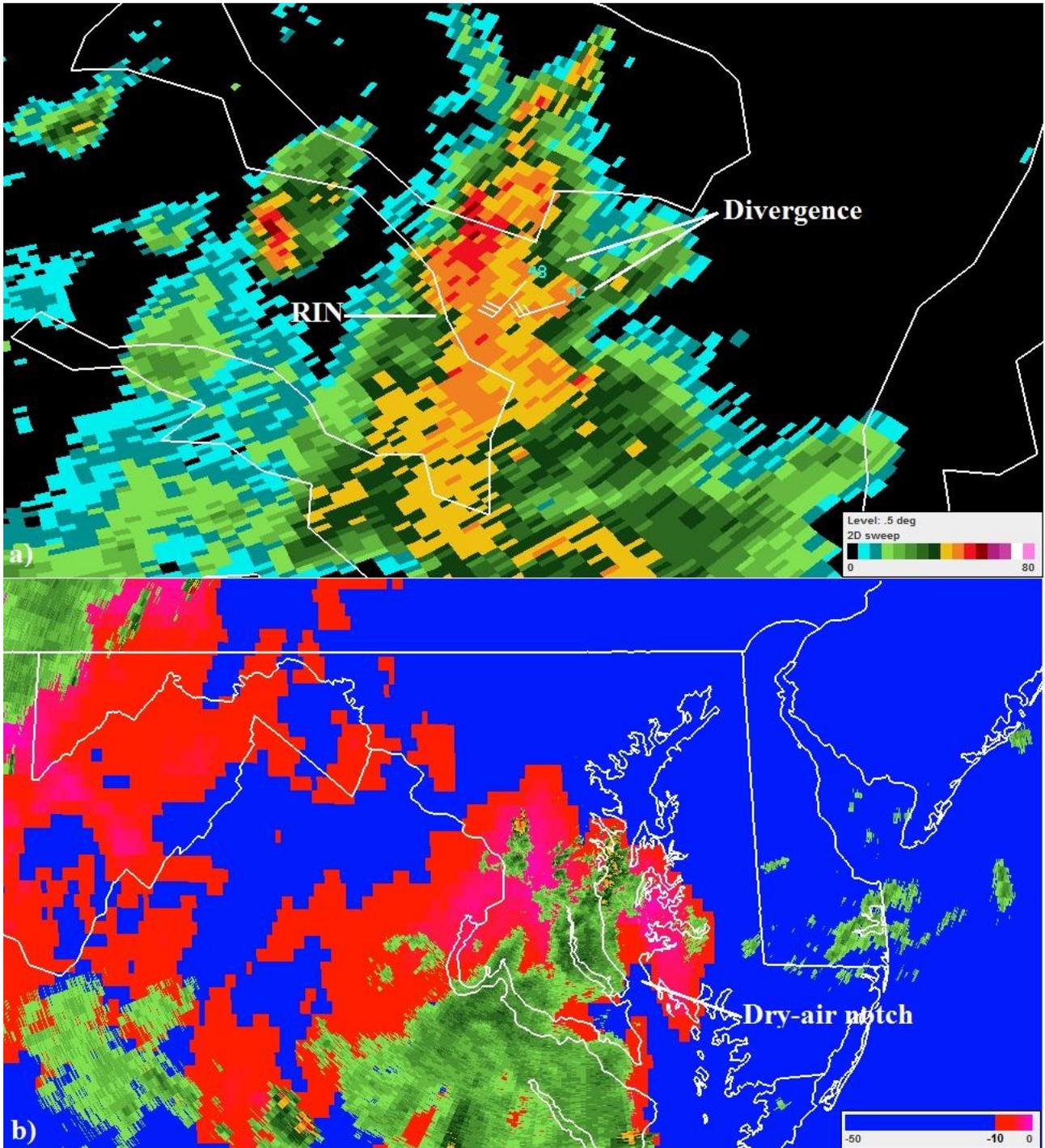

Figure 7. a) Baltimore, Maryland TDWR reflectivity (dBZ) at 0029 UTC 11 July 2013 with overlying wind barbs at Greenbury Point Weatherflow station and the NOAA Annapolis Buoy and b) GOES-13 WV-IR brightness temperature difference (BTD) at 0015 UTC 11 July 2014 with overlying TDWR reflectivity at 0029 UTC. "48" and "42" represent wind gusts recorded by the Greenbury Point station and by the Annapolis Buoy, respectively, near 0030 UTC. NEXRAD contour interval is 5 dBZ. In b), blue shading indicates BTD < -10°K associated with mid-tropospheric subsaturated air while magenta shading indicates BTD > 0°K associated with cold thunderstorm cloud tops.

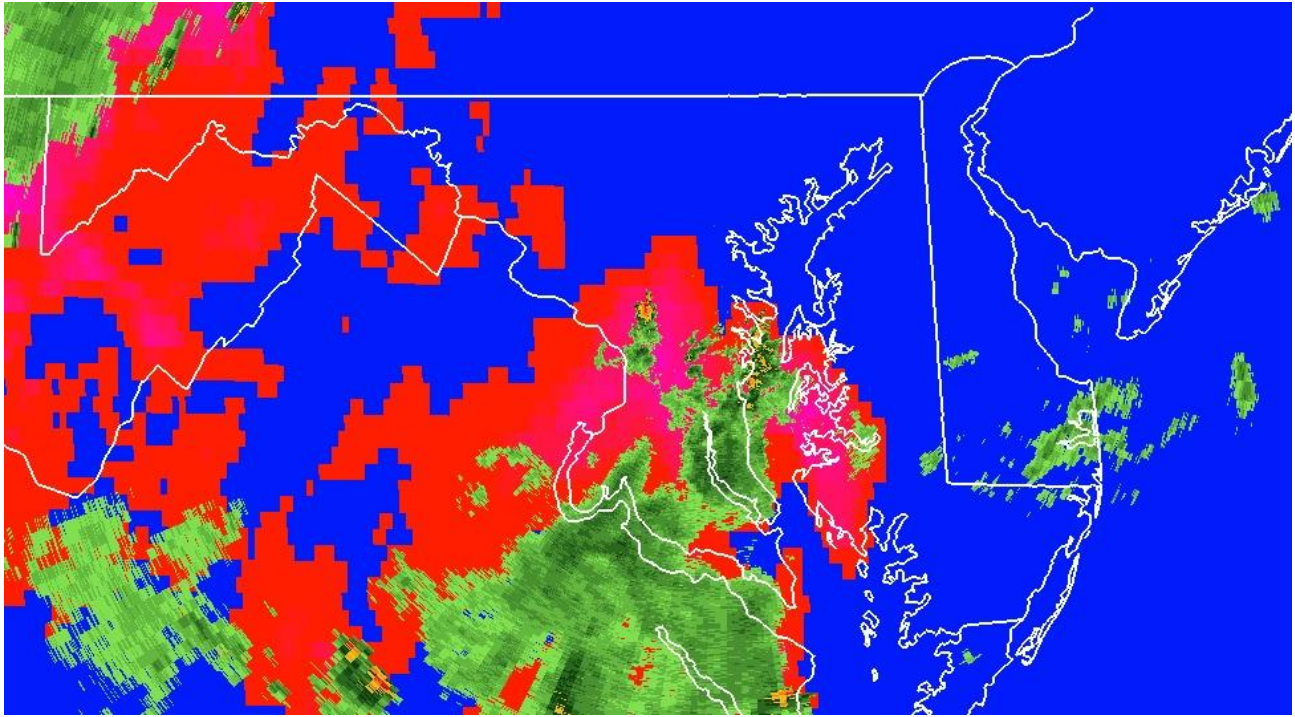

Video 1. A composite GOES BTD/TDWR reflectivity image animation between 0015 and 0045 11 July 2013. http://dx.doi.org/doi.number.goes.here